\documentclass{iopart}

\usepackage{epsfig}
\usepackage{psfig}

\newcommand{\Z}{{\mathcal Z}}

\newcommand{\Ze}{{\mathcal Z}_{\rm ex}}

\newcommand{\kb}{k_{\rm B}}

\newcommand{\ef}{\epsilon_{\rm F}}

\begin{document}

\jl{1}
\title[Fluctuations of the Fermi condensate]{Fluctuations of the Fermi condensate in ideal gases}

\author{Drago\c s-Victor Anghel,$^{1,2}$\footnote{E-mail: dragos@fys.uio.no}, Oleg Fefelov$^{1}$\footnote{E-mail: olegfe@fys.uio.no} and Y. M. Galperin$^{1}$\footnote{E-mail: yurig@fys.uio.no}}

\address{$^1$University of Oslo, Department of Physics, P.O.Box 1048
Blindern, N-0316 Oslo, Norway,\\ $^2$NIPNE  -- ``HH'', P.O.Box MG-6,
R.O.-76900 Bucure\c sti - M\u agurele, Rom\^ania}

\begin{abstract}
We calculate numerically and analytically the fluctuations of the fermionic 
condensate and of the number of particles above the condensate for systems of
constant density of states. We compare 
the canonical fluctuations, obtained from the equivalent Bose condensate
fluctuation, with the grandcanonical fermionic calculation. The
fluctuations of the condensate are almost the same in the two ensembles, 
with a small correction comming from the total particle number 
fluctuation in the grandcanonical ensemble. On the other hand the number 
of particles above the condensate and its fluctuation is insensitive 
to the choice of ensemble. 
\end{abstract}
\pacno{05.30.Ch, 05.40.-a, 05.90.+m}
\submitto{\JPA}
\maketitle

\section{Introduction}\label{intro1}

Starting quite a long time ago, Auluck and Kothari \cite{auluck}, then May 
\cite{may} and finally Viefers, Ravndal, and Haugset \cite{viefers}, 
discovered idependently that the specific heat of nonrelativistic ideal 
gases in two-dimensional (2D) boxes does not depend on the exclusion 
statistics. This interesting result eventually did not 
receive the attention it deserved until 1995, when Lee \cite{lee} rederived it 
by introducing an unified way of writing the thermodynamic properties of 
ideal gases 
in terms of polylogarithmic functions \cite{lewin}. This formulation 
represented also an important extension of the Auluck and Kothari result 
and triggered further investigations (see for example Refs. 
\cite{apostol,pathria,paper1}). 

Since under {\em canonical conditions} a Bose and its corresponding Fermi gas 
are similar at the thermodynamic level, they have been called 
{\em thermodynamically equivalent}. If we denote by $C_V(T,V,N)$ the 
heat capacity of a system at temperature $T$, volume $V$ 
and particle number $N$, then the heat capacities $C_{V,1}$ and 
$C_{V,2}$ of two thermodnamically equivalent systems are identical 
functions of $T$, $V$ and $N$. Using this property, all the thermodynamic 
systems may be divided into equivalence classes \cite{paper1} 
and by doing this one may observe that all the systems of ideal particles 
of the same constant density of states (DOS), but obeying Bose, Fermi, or even 
{\em fractional exclusion statistics} \cite{haldane}, belong to the same 
equivalence class \cite{paper1}. 

The equivalence between Bose and Fermi gases was critically examined 
by Pathria \cite{pathria}. 
He showed that the Lee's unified formulation of 2D ideal gases 
does not hold anymore below the Bose-Einstein condensation temperature 
of the Bose gas. Aparently, the 2D (or, more exactly, constant DOS) 
{\em thermodynamic equivalence holds only above the Bose-Einstein condensation temperature}. 

On the other hand, Crescimanno and Landsberg \cite{crescimanno} and one of 
us \cite{paper1} showed that there is a one-to-one mapping 
between microscopic configurations of bosons, fermions or haldons, in 
systems with the same, constant DOS, which preserves the total excitation 
energy (i.e. the energy of the particles in the given configuration minus 
the energy of the system at zero kelvins is the same). Based on this 
theorem, the thermodynamic equivalence of systems with equally spaced 
spectra {\em should hold at any temperature in any detail}, so 
Pathria's conclusion must be wrong. But what was overlooked there? 

The method of mapping microscopic configurations between systems of 
different exclusion statistics introduced in 
Ref. \cite{paper1} for systems with constant DOS was extended in Ref. 
\cite{paper3} for systems with any DOS and was called 
{\em exclusion statistics transformation} (EST). Systems connected by 
EST are thermodynamically equivalent by construction. If we take for 
example a Fermi system, transform it by EST into a Bose system, and 
then calculate independently the thermodynamical properties  of these 
two systems by maximizing the entropy in each of them, for constant 
$U$ (total energy) and $N$ -- i.e. assuming grandcanonical distribution 
on the single particle energy levels -- we loose again the thermodynamic 
equivalence that we started with, at least is most of the cases 
\cite{paper3}! The obvious conclusion 
is that one or both of these grandcanonical distributions lead to results 
in dezacord with the canonical ensemble. The question which of these 
grandcanonical distribution is closest to the canonical distribution is 
very dificult to answer, since {\em ab initio} canonical calculations 
are not easy to perform on general, large systems. 

%In this paper we shall analyse systems with constant DOS, which are 
%easiest to handle and provide a very clear example. If the spectrum 
%is assumed to consist of equidistant single particle levels, then by 
%EST the ideal Fermi gas is transformed into an ideal Bose gas of identical 
%spectrum and {\em vice-versa}. In this system both, grandcanonical Bose 
%distribution and grandcanonical Fermi distribution describe 
%very well the unified canonical distribution, provided that the 
%{\em fermionic condensation} \cite{paper2} is taken into acount 
%\cite{paper4}. 

%The fermionic condensate -- the lowest $N_0$ single 
%particle levels, ocupied with probability 1 -- is put into correspondence 
%with the Bose condensate. To make the paper easier to read, in Section 
%\ref{intro}
%we review briefly the concept of Fermi condensate. In the following 
%sections we calculate in detail and compare the ...

\subsection{The Fermi condensate}
\label{intro}

The concept of Fermi condensate was introduced in Refs. \cite{paper3,paper2}. 
For a general system, of spectrum consisting of single particle energy 
levels $\epsilon_i$ (we enumerate them such that 
$\epsilon_0\le\epsilon_1\le\ldots$), one can calculate the
grandcanonical probability, $w_{N_0}$, of having the lowest $N_0$ energy
levels occupied, the level $N_0+1$ free and all the other energy levels with
any occupation number \cite{paper2}:
\begin{equation} \label{wN0}
\fl
w_{N_0} = \Z^{-1}\cdot\exp\left[-\beta\cdot\sum_{i=0}^{N_0-1}\epsilon_i
+\beta\mu N_0\right]\cdot\Ze(N_0,\beta,\beta\mu)
\equiv \frac{\Z_{N_0}}{\Z} \, .
\end{equation}
Here $\Z$ is the partition function of the system and 
$\Ze(N_0,\beta,\beta\mu)=\prod_{i=N_0+1}^\infty\left\{1+\exp[-\beta(\epsilon_i-\mu)]\right\}$
is the partition function of the levels $N_0+1, N_0+2,\ldots$. Obviously,
\begin{equation} \label{Ztot}
\Z = \sum_{N_0} \Z_{N_0} \,.
\end{equation}
The probability distribution (\ref{wN0}) may have a maximum at, say
$N_{\rm 0,max}$. The statistical interpretation of such a maximum is
that in a physical system in contact with a particle reservoir, the
lowest $N_0(\approx N_{\rm 0,max})$ energy levels are always occupied.
These $N_0$ particles form the {\em Fermi condensate}. 
At any finite temperature, $N_0$ is subject to fluctuations, denoted 
by $\delta N_0$. 

The configurations of fermions may be transformed by EST into configurations 
of bosons \cite{paper3,paper1}. By this transformation the $N_0$ degenerate 
fermions will be transformed into the Bose-Einstein condensate of the 
corresponding Bose system, and hence the name of Fermi condensate. 
For this reason the degenerate fermions will also be called 
Fermi condensate. For canonical Bose systems of constant DOS, the 
probability distribution of having 
$N_0$ particles in the condensate, $w^{\rm c}_{N_0}$, have
been studied in detail before (see for example Ref.\cite{mullin} for a
review). By construction, the canonical probability distribution of
having $N_0$ fermions in the condensate is also $w^{\rm c}_{N_0}$. 

In Ref. \cite{paper4} it was shown that in a system of constant density
of states, $N_{\rm 0,max}$ -- which corresponds to the maximum of $w_{N_0}$ 
(\ref{wN0}) --
is close to the average canonical ocupation of the ground state, but the
two numbers are not the same. The distribution $w_{N_0}$ is not symmetric 
around the maximum, so $\langle N_0\rangle\ne N_{\rm 0,max}$ (where 
by $\langle\cdot\rangle$ and $\langle\cdot\rangle^{\rm c}$ we shall denote 
grandcanonical and canonical averages, respectively). 
Since if the Bose and Fermi condensates 
are separated from the rest of the particles, the grandcanonical 
Bose and Fermi distributions map onto each-other (see Section 5, Ref. 
\cite{paper3}), the average number of particles in the 
Fermi condensate should be equal to the average ocupation of 
the bosonic ground state and in the Fermi system below the 
condensation temperature the usual Fermi distribution applies only to 
the particles above the condensate. 
For simplicity, to the fermions above the condensate 
we shall refer to as the particles in the {\em thermally active layer}. 

It is well known that Fermi canonical and grandcanonical ensembles are 
equivalent. In this paper we shall investigate the equivalence between 
the grandcanonical description of the Fermi gas and its correspondent 
canonical description in terms of the Bose gas. This is a new type of 
equivalence, first mentioned in Ref. \cite{paper3} and which seems not to 
hold for general systems. Here we shall discuss the simplest systems, 
namely the ones with constant density of states, $\sigma$, and 
we shall compare $w_{N_0}$ and $w^{\rm c}_{N_0}$. An important 
parameter of the system is the quantity $\sigma\kb T$. For large 
$\sigma\kb T$ we can do some analytical calculations, 
assuming that $w_{N_0}$ has a gaussian shape, but, as we shall see in 
seection \ref{analytical}, this approximation is not too good for the 
evaluation of $\langle\delta^2 N_0\rangle$, the mean square fluctuation 
of $N_0$. To correct this and to extend our calculations to lower values of 
$\sigma\kb T$, in Section \ref{exact} we do numerical 
calculations of $w_{N_0}$, $w^{\rm c}_{N_0}$, 
$\langle\delta^2 N_0\rangle$ and $\langle\delta^2 N_0\rangle^{\rm c}$. 
(For simplicity we use the somewhat simpler notations 
$\langle\delta^2 N_0\rangle$ and $\langle\delta^2 N_0\rangle^{\rm c}$ 
instead of the more rigorous $\langle(\delta N_0)^2\rangle$ and 
$\langle(\delta N_0)^2\rangle^{\rm c}$, respectively.)
We reobtain the known result: 
$\langle\delta^2 N_0\rangle^{\rm c}\to\zeta(2)(\sigma\kb T)^2$, i.e. the 
mean square fluctuation of $N_0$ is of the same order as the number of 
particles in the thermally active layer or the number of particles on 
the excited states in the Bose gas. We also obtain 
that $\langle\delta^2 N_0\rangle-\langle\delta^2 N_0\rangle^{\rm c}$ 
converge to a constant value, 0.39, as $\sigma\kb T$ increases. 
This asymptotic behavior is proved analytically by the end of 
Section \ref{exact}. 

\section{Analytical evaluation of the fluctuations}
\label{analytical}

\begin{figure}[t]
\begin{center}
\unitlength1mm\begin{picture}(60,65)(0,0)
\put(0,0){\epsfig{file=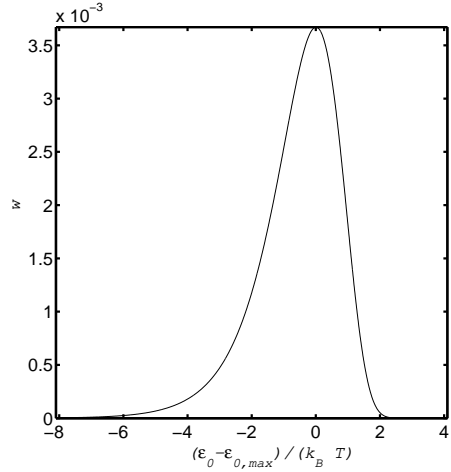,width=60mm}}
\end{picture}
\caption{Probability distribution of $w_{N_0}$ as a function of
$N_0$. The maximum of the probability distribution is located at $N_{0,max}$, 
which is given by the equation 
$(\mu-N_{0,max}/\sigma)/k_BT=\log(\sigma k_BT)$. 
For this particular plot $\sigma\kb T=100$.}
\label{probability}
\end{center}
\end{figure}

First we analyse Eq. (\ref{wN0}) analytically in the limit
$\sigma\kb T\gg 1$. To do this we take $\log w_{N_0}$, and
transform all the summations into integrals. In this way we arrive to the
expression \cite{paper2}
\begin{eqnarray}
\fl
\log\Z_{N_0} &=& \left[ -\beta \left(\sigma\frac{\epsilon_0^{2}}{2}-
\epsilon_0\right)+\beta\mu \left(\sigma\epsilon_0-1\right) \right]
+ \sigma\int_{\epsilon_0}^\infty d\epsilon\,
\log\left[1+\e^{-\beta(\epsilon-\mu)} \right] \, ,\label{logZN0}
\end{eqnarray}
where $\epsilon_0$ is the energy of the $N_0^{th}$ single-particle level, the integral represents $\log \Ze(N_0,\beta,\beta\mu)$.
The shape of the probability distribution is depicted in figure
\ref{probability}.
Since $\sigma\epsilon_0=N_0$ and $\partial\log\Z_{N_0}/\partial N_0=\sigma^{-1}(\partial\log\Z_{N_0}/\partial\epsilon_0)$, the value of $\epsilon_0$
corresponding to the maximum of probability is given by the equation 
\begin{eqnarray}
\fl
\left.\frac{\partial\log \Z_{N_0}}{\partial \epsilon_0}\right|_{
\epsilon_{\rm 0,max}} &=& 
-\sigma\left\{\log\left[1+\e^{\beta(\epsilon_{\rm 0,max}-\mu)} \right]
- (\sigma\kb T)^{-1}\right\} = 0\, , \label{HA!}
\end{eqnarray}
and for $\sigma\kb T\gg 1$ \cite{paper4} 
\begin{equation} \label{epsilon02d}
\epsilon_{\rm 0,max} = \mu -k_B T\log[\sigma k_B T] \, .
\end{equation}
We observe here that $\beta(\mu-\epsilon_{\rm 0,max})$ depends only on 
$\sigma k_B T$ and not on $\mu$, i.e. on the total particle number,  
as long as $\mu>\kb T\log[\sigma k_B T]$. Therefore, as $T$ decreases 
and $\mu$ becomes bigger than $\kb T\log[\sigma k_B T]$ the probability 
distribution (\ref{wN0}) forms a maximum at $N_0>0$. We say that at this 
temperature the condensate starts to form and the equation 
$\mu=k_B T\log[\sigma k_B T]$ defines the condensation temperature. 

At low temperatures the maximum of $w_{N_0}$ becomes sharp and 
$\epsilon_0$ approaches $\mu$. In this temperature range we shall
approximate $w_{N_0}$ around the maximum by a gaussian distribution: 
\begin{equation}
w_{N_0}\approx w(N_{\rm 0,max})\cdot\exp\left[-\frac{(N_0-N_{\rm 0,max})^2} 
{2\Delta^2}\right] \label{gaussappr}
\end{equation}
The width of the gaussian is 
\begin{equation}
\fl
\Delta^{-2} = -\left.\frac{\partial^2\log \Z_{N_0}}{\partial N_0^2}\right|
_{N_{\rm 0,max}} =(\sigma\kb T)^{-1}-(1+\sigma\kb T)^{-1}\approx
(\sigma\kb T)^{-2} \,. \label{deltaG}
\end{equation}
Equation (\ref{gaussappr}) amounts to the use of the approximative 
function $\Z^{\rm (a)}_{N_0}$ instead of $\Z_{N_0}$: 
\begin{equation}
\fl
\Z^{\rm (a)}_{N_0} = \exp\left\{\log \Z_{\tilde N_0}+\frac{1}{2}\cdot
\frac{\partial^2\log \Z_{N_0}}{\partial N_0^2} \delta^2 N_0 \right\} 
= \Z_{N_{\rm 0,max}}\exp\left\{-\frac{1}{2}\cdot
\frac{\delta^2 N_0}{(C\kb T)^2} \right\} \, . 
\end{equation}
To check the approximation we calculate first the total partition function 
(\ref{Ztot}) as 
\begin{eqnarray}
%\fl
\Z^{\rm (a)} &=& \int_{-\infty}^\infty \rmd (\delta N_0) \Z^{\rm (a)}_{N_0} 
\approx \Z_{N_{\rm 0,max}}\cdot\int_{-\infty}^\infty \rmd (\delta N_0) 
\exp\left[-\frac{1}{2}\cdot\frac{\delta^2 N_0}{(\sigma\kb T)^2} \right]
\nonumber \\
&=& Z_{N_{\rm 0,max}} \sqrt{2\pi} \sigma\kb T \, , \label{Za}
\end{eqnarray}
where
\begin{eqnarray} 
\fl
\log\Z_{N_{\rm 0,max}} &=& -\beta \left(\sigma\frac{\epsilon_{\rm 0,max}^{2}}
{2}-\epsilon_{\rm 0,max}\right)+\beta\mu \left(\sigma\epsilon_{\rm 0,max} 
-1\right) + \sigma\int_{\epsilon_{\rm 0,max}}^\infty \rmd\epsilon\, 
\log\left[1+\e^{-\beta(\epsilon-\mu)} \right] \nonumber \\
\fl
&=& \frac{\sigma\kb T}{2}\left[(\beta\mu)^2-\log^2(\sigma\kb T) \right] 
-\log(\sigma\kb T) + \sigma\kb T Li_2\{-\exp[\beta(\mu-\epsilon_{\rm 0,max})
]\} \, , \nonumber \\ \fl &&\label{logZtN0}
\end{eqnarray}
where $Li_2$ is Euler's dilogarithm function \cite{lewin}. 
Using Eq. (\ref{epsilon02d}) and the expansion 
\begin{equation}
\left.Li_2(z)\right|_{z\ll -1}\approx\frac{(\log|z|)^2}{2}
\left[1+\frac{\pi^2}{6}\frac{2}{(\log|z|)^2}\right] \,, \label{Li2approx}
\end{equation}
valid for any $n>1$, we obtain the approximation
\begin{equation}
\log\Z^{\rm (a)} \approx \frac{\sigma\kb T}{2}(\beta\mu)^2 + 
\frac{\pi^2}{6}\sigma\kb T + \log\sqrt{2\pi} \, . \label{ZalowT1}
\end{equation}

On the other hand, the exact partition function is 
\begin{eqnarray} 
\log\Z &=& \sigma\int_{0}^\infty \rmd\epsilon\, 
\log\left[1+\e^{-\beta(\epsilon-\mu)} \right] =\sigma\kb T Li_2[-\exp(
\beta\mu)] \,,
\label{Zexact}
\end{eqnarray}
which, if we apply again the approximation (\ref{Li2approx}) we get 
\begin{eqnarray} 
\log\Z &=& \frac{\sigma\kb T}{2} (\beta\mu)^2 + \frac{\pi^2}{6} 
\sigma\kb T \, .\label{Zexact1}
\end{eqnarray}
The expansions of $\log\Z$ and $\log\Z^{\rm (a)}$ are identical 
up to order $\sigma\kb T/(\beta\mu)^2$, or $\sigma\kb T/\log^2(C\kb T)$. 
The term $\log\sqrt{2\pi}\approx 0.92$ from equation (\ref{ZalowT1}) may be 
neglected, since is smaller than the order 
$\sigma\kb T/\log^2(\sigma\kb T)\gg 1$, for $\sigma\kb T\gg 1$. 

The fluctuation of the number of particles in the condensate, in the 
approximation (\ref{gaussappr}), is 
\begin{equation} \label{FdeltaN}
\sqrt{\langle \delta^2 N_0\rangle} = \sigma \kb T \, .
\end{equation}
On the other hand, the corresponding canonical fluctuation may be 
calculated for example by saddle point method applied to the equivalent 
Bose system \cite{mullin,holthaus} and gives 
\begin{equation} \label{FdeltaNc}
\sqrt{\langle \delta^2 N_0\rangle^{\rm c}} = \sqrt{\zeta(2)}\sigma \kb T \,,
\end{equation}
where $\zeta(x)$ is the Riemann Zeta function. Obviously, the two analytical
approximations, (\ref{FdeltaN}) and (\ref{FdeltaNc}), do not coincide and 
the question that remains to be answered is whether these distributions
are indeed different, or simply the gaussian approximation (\ref{gaussappr}) 
is not good enough. 

\section{Numerical evaluation of the fluctuations}
\label{exact}

In this section we calculate 
the fluctuations numerically by introducing a recursion relation. 
From Eq. (\ref{wN0}) we obtain
\begin{equation}
\frac{w_{N_0+1}}{w_{N_0}} = \frac{\exp[-\beta(\epsilon_{N_0}-\mu)]}
{1+\exp[-\beta(\epsilon_{N_0+1}-\mu)]} \label{recursion}
\end{equation}
and the value of $N_{\rm 0,max}$ may be found by solving
\begin{equation}
\frac{\exp[-\beta(\epsilon_{N_{\rm 0,max}}-\mu)]}
{1+\exp[-\beta(\epsilon_{N_{\rm 0,max}+1}-\mu)]} = 1\,. \label{Nmaxnum}
\end{equation}
If the density of states is constant and 
$\epsilon_{i+1}-\epsilon_i=\sigma^{-1}$ for any $i$, 
then Eq. (\ref{Nmaxnum}) becomes 
\begin{equation}
\frac{\exp[-\beta(\epsilon_{\rm 0,max}-\mu)]}
{1+\exp[-\beta(\epsilon_{\rm 0,max}+\sigma^{-1}-\mu)]} = 1\,. \label{Nmaxnumc}
\end{equation}
Using Eqs. (\ref{recursion}) and (\ref{Nmaxnumc}) we may now calculate 
numerically $N_{\rm o,max}$, $\langle N_0\rangle$, and 
$\langle \delta^2 N_0\rangle$. If $\sigma\kb T\gg1$, by writing 
$\exp[-\beta(\epsilon_{\rm 0,max}+\sigma^{-1}-\mu)]\approx\exp[-\beta(\epsilon_{\rm 0,max}-\mu)](1-(\sigma\kb T)^{-1})$ Eq. (\ref{Nmaxnumc}) may be 
simplified to 
$\exp[\beta(\mu-\epsilon_{\rm 0,max})] = \sigma\kb T$, 
which is the same as Eq. (\ref{epsilon02d}). Moreover, since around the 
maximum $\exp[\beta(\mu-\epsilon_{\rm 0,max})]\gg 1$, in the relevant 
energy interval we may transform Eq. (\ref{recursion}) into 
\begin{equation}
\fl
\frac{w_{N_0+1}}{w_{N_0}} = \left\{\exp[\beta(\epsilon_{N_0}-\mu)]+
\exp[-(\sigma\kb T)^{-1}]\right\}^{-1} \approx 1-\e^{\beta(\epsilon_{N_0}-\mu)}
+(\sigma\kb T)^{-1} \,.
\label{recursionlim}
\end{equation}

Let us now analyse the equivalent Bose gas. If again the system has a 
constant density of states we 
denote $q\equiv\e^{-1/(\sigma\kb T)}$. Then the canonical 
partition function for a system of $N_{\rm ex}$ particles is \cite{mullin}
\begin{equation}
Z^{\rm b}_{N_{\rm ex}} = \prod_{k=1}^{N_{\rm ex}} \frac{1}{1-q^k} \,. 
\label{mullin}
\end{equation}
In a canonical system of $N$ particles, the probability 
$w^{\rm b}_{N_{\rm ex}}$ to have exactly $N_{\rm ex}$ particles in the 
excited states (not on the ground state) is proportional to 
$Z^{\rm b}_{N_{\rm ex}}-Z^{\rm b}_{N_{\rm ex}-1}$ \cite{holthaus}, so we have 
\begin{equation}
w^{\rm b}_{N_{\rm ex}} = \frac{q^{N_{\rm ex}}}{Z}
\prod_{k=1}^{N_{\rm ex}} \frac{1}{1-q^k} \,. \label{probabNex}
\end{equation}
Since $N_{\rm ex}\equiv N-N_0$, the relative probability which corresponds 
to Eq. (\ref{recursion}) for fermions is 
\begin{equation}
\frac{w^{\rm b}_{N_{\rm ex}-1}}{w^{\rm b}_{N_{\rm ex}}} = \frac{1-q^{N_{\rm ex}}}{q} = \left\{1-\exp[N_{\rm ex}/(\sigma\kb T)]\right\}\exp[(\sigma\kb T)^{-1}]\,. \label{recursionB}
\end{equation}
The most probable $N_{\rm ex}$ is given by 
\begin{equation}
\left[1-\exp(-\beta N_{\rm ex}/\sigma)\right]\exp[(\sigma\kb T)^{-1}] = 1 \,.
\label{NmaxnumcB}
\end{equation}

We want now to compare Eqs. (\ref{recursion}) and (\ref{recursionB})
in the limit $\sigma\kb T\gg 1$. For this
we take a Fermi and a Bose system with the same number of particles, $N$.
In the Fermi system we define $\ef=N/\sigma$ (Fermi energy).
We shall assume that both systems are below the condensation temperature and
the number of particles in the condensate is $N_0$. Above the
condensate we have $N_{\rm ex}$ particles. For a condensed gas
$\ef-\mu<\sigma^{-1}$, so we can express $N_{\rm ex}$ in Eq.
(\ref{recursionB}) as
$N_{\rm ex}=\sigma(\ef-\epsilon_0)=\sigma(\mu-\epsilon_0)$. By doing
so, Eq. (\ref{recursionB}) becomes
\begin{eqnarray}
\frac{w^{\rm b}_{N_0+1}}{w^{\rm b}_{N_0}} &=& \left\{1-\exp[\beta(
\epsilon_{n_0}-\mu)]\right\}\exp[(\sigma\kb T)^{-1}] \nonumber \\
&\approx& 1-\exp[\beta(\epsilon_{n_0}-\mu)] + (\sigma\kb T)^{-1}
\,, \label{recursionBlim}
\end{eqnarray}
which is identical to Eq. (\ref{recursionlim}). Therefore the
two probability distributions $w_{N_0}$ and $w^{\rm b}_{N_0}$ approach 
each-other in the limit of large systems, i.e. when $\sigma\kb T\gg 1$.
%nd both represent the probability of having $N_0$ particles condensed.

\begin{figure}[t]
\begin{center}
\unitlength1mm\begin{picture}(100,55)(0,0)
\put(0,0){\epsfig{file=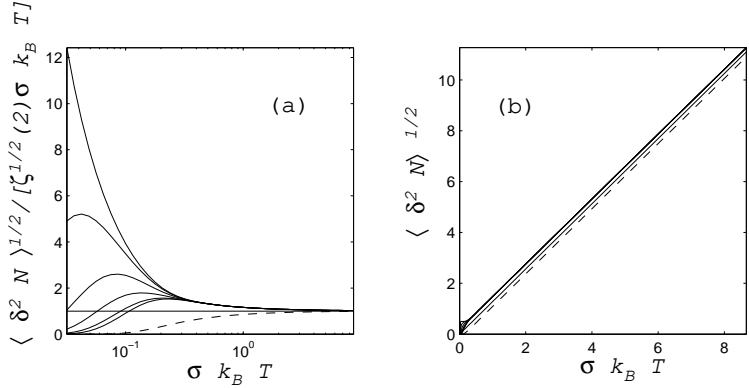,width=100mm}}
\end{picture}
\caption{Numerical calculations of
$\langle\delta^2 N\rangle^{1/2}/[\zeta^{1/2}(2)\sigma\kb T]$ (a) and
$\langle\delta^2 N\rangle^{1/2}$ (b). For low values of $\sigma\kb T$,
$\langle\delta^2 N\rangle$ depends on the exact
location of $\mu$, with respect to the single particle levels. If
$\epsilon_N$ and $\epsilon_{N+1}$ are two consecutive energy levels
($\epsilon_{N+1}-\epsilon_N=\sigma^{-1}$), then the solid curves in
both (a) and (b) figures correspond to
$\mu=\epsilon_N+0.1\cdot i\sigma^{-1}$ ($i=0,1,\ldots,5$) from top to bottom. The
dashed lines represent the canonical fluctuations, and the straight lines
($y=1$ in a and $y=\zeta^{1/2}(2)x$ in b) correspond to the
asymptotic values.}
\label{fluctuationsM}
\end{center}
\end{figure}

The numerical calculations, based on Eqs. (\ref{recursion}) and
(\ref{recursionB})
are plotted in figure \ref{fluctuationsM}. We can observe that already for
$\sigma\kb T$ bigger than 1, the fluctuation of the particle number in the
condensate is almost the same for the canonical Bose and grandcanonical
Fermi systems. This justify the approach taken in Ref. \cite{paper4},
and for $\sigma\kb T\gg 1$, $N_0$ may be calculated directly as the average
number of particles in the Bose condensate, rather than by Eq. (\ref{HA!}).

Noticeble {\em relative} differences between canonical and grandcanonical 
results appear only for $\sigma\kb T$ about 1 or below.
For these values of $\sigma\kb T$ the fluctuations depend on the exact
location of $\mu$, with respect to the single particle levels.
For example let's say that $\mu\in(\epsilon_{N-1},\epsilon_{N})$,
where $\epsilon_{N-1}$ and $\epsilon_{N}$ are two consecutive energy levels.
In the limit $\beta(\mu-\epsilon_{N-1})\to\infty$ and for $N_0=N$, equation
(\ref{recursion}) becomes
\begin{equation}
\frac{w_{N}}{w_{N-1}} \approx \exp[\beta(\mu-\epsilon_{N-1})] \,.
\label{recursionlim1}
\end{equation}
For $\beta(\mu-\epsilon_{N-1})\to\infty$ we can calculate $\langle N_0\rangle$
and $\langle\delta^2 N_0\rangle$
by taking into account only the levels $\epsilon_{N-1}$ and $\epsilon_{N}$
and we obtain
\begin{equation}
\langle N_0\rangle = \frac{\exp[\beta(\mu-\epsilon_{N-1})](N+1)+N}
{\exp[\beta(\mu-\epsilon_{N-1})]+1} = N+ 1 - \exp[-\beta(\mu-\epsilon_{N-1})]
\end{equation}
and
\begin{eqnarray}
\fl
\langle\delta^2 N_0\rangle &=& \frac{\exp[\beta(\mu-\epsilon_{N-1})]
\exp[-2\beta(\mu-\epsilon_{N-1})]+(1-\exp[-\beta(\mu-\epsilon_{N-1})])^2}
{\exp[\beta(\mu-\epsilon_{N-1})]+1} \nonumber \\
\fl
&=& \frac{1-\exp[-\beta(\mu-\epsilon_{N-1})]}{\exp[\beta(\mu-\epsilon_{N-1})]
+1} \approx \exp[-\beta(\mu-\epsilon_{N-1})] \label{delta2Nnum}
\end{eqnarray}
Therefore, for any $\mu\in(\epsilon_N,\epsilon_{N+1})$, 
\begin{equation}
\lim_{T\to 0}\left\{\langle\delta^2 N_0\rangle^{1/2}\left[\zeta^{1/2}(2)\sigma\kb T\right]^{-1}\right\}=0 \,. \label{limT0int}
\end{equation}

The situation is different if for example $\mu=\epsilon_N$. Then,
applying the same algorithm as above, we get $\langle N_0\rangle=N+0.5$
and $\langle\delta^2 N\rangle^{1/2}=0.5$. In figure 2 we plotted
$\langle\delta^2 N\rangle^{1/2}/[\zeta^{1/2}(2)\sigma\kb T]$ (a) and
$\langle\delta^2 N\rangle^{1/2}$ (b) for $\mu=\epsilon_N+0.1\cdot i\sigma^{-1}$
($i=0,1,\ldots,5$). The fluctuations normalized to the asymptotic
value, $\zeta^{1/2}(2)\sigma\kb T$, are quite different for
$\sigma\kb T\le 1$, but the absolute values of the fluctuations are
very close for any $\sigma\kb T$ for both types of systems and 
any choice of $\mu$.

We notice also in figure \ref{fluctuationsM} (b) that although
the difference
$\sqrt{\langle\delta^2 N_0\rangle}-\sqrt{\langle\delta^2 N_0\rangle^{\rm c}}$
is very small for any $\sigma\kb T$, it does not converge
to zero as $\sigma\kb T\to\infty$. Numerically we obtain 
\begin{equation}
\sqrt{\langle\delta^2 N_0\rangle}-\sqrt{\langle\delta^2 N_0\rangle^{\rm c}} 
\approx 0.39 \qquad {\rm for}\ \sigma\kb t\gg 1 \,. \label{fluctas1}
\end{equation}
To explain this difference, let us note that the fluctuation of 
$N_0$ in the grandcanonical ensemble, 
$\delta N_0\equiv N_0-\langle N_0\rangle$, may be viewed as the superposition 
of fluctuations comming from two sources: the canonical fluctuation of 
$N_0$ around its average value, corresponding to the total particle number 
$N$, denoted by $\delta^N N_0 (\equiv N_0-\langle N_0\rangle_N)$, 
and the fluctuation of 
$\langle N_0\rangle_N$ due to the grandcanonical fluctuation of $N$. 
Assuming small fluctuations, the variation of $\langle N_0\rangle_N$ 
due to the variation of $N$ may be written as 
\[
\delta\langle N_0\rangle_N = \frac{\partial \langle N_0\rangle_N}{\partial 
N}\cdot\delta N\,.
\]
Collecting all these together we write 
\begin{equation}
\fl
\delta N_0 = N_0-\langle N_0\rangle \equiv N_0-\langle N_0\rangle_N + 
\langle N_0\rangle_N - \langle N_0\rangle = \delta N_0^{\rm c} + 
\frac{\partial \langle N_0\rangle}{\partial N} \delta N \,. 
\label{deltaN0s}
\end{equation} 
Below the condensation, $\partial \langle N_0\rangle/\partial N=1$ 
(temperature stays constant). Moreover, well below the 
condensation temperature $\delta N_0^{\rm c}$ and $\delta N$ are 
independent, since the condensate may be viewed as a reservoir of particles of 
zero energy \cite{maxwell}, and Eq. (\ref{deltaN0s}) leads to  
\begin{equation}
\langle\delta^2 N_0\rangle = \langle\delta^2 N_0\rangle^c + 
\langle\delta^2 N\rangle \,. \label{squarefl1} 
\end{equation}
For high enough $\sigma\kb T$ and $\beta\mu$ we have 
$\langle\delta^2 N\rangle\approx\sigma\kb T$, which, if plugged in 
Eq. (\ref{squarefl1}) gives 
\begin{equation}
\langle\delta^2 N_0\rangle\approx\sqrt{\zeta(2)(\sigma\kb T)^2+\sigma\kb T} 
\approx\sqrt{\zeta(2)}\sigma\kb T+\frac{1}{2\sqrt{\zeta(2)}} \,. 
\label{evrika}
\end{equation}
As we expect, $(2\sqrt{\zeta(2)})^{-1}\approx 0.39$. 

Using the same method we calculate the fluctuation of the number 
of particles in the thermally active layer without doing any extra 
numerics. Again we denote by $\langle N_{\rm ex}\rangle$ the average 
number of particles in the thermally active layer and the fluctuation 
$\delta N_{\rm ex}$ can again be written as 
\begin{equation}
\fl
\delta N_{\rm ex} \equiv N_{\rm ex}-\langle N_{\rm ex}\rangle = 
N_{\rm ex} - \langle N_{\rm ex}\rangle_N + \langle N_{\rm ex}\rangle_N -
\langle N_{\rm ex}\rangle = \delta^{\rm c}N_{\rm ex} + 
\frac{\partial \langle N_{\rm ex}\rangle}{\partial N} \delta N \,.
\label{fluctNex}
\end{equation}
By $\langle N_{\rm ex}\rangle_N$ we denote the average number of particles 
in the thermally active layer at fixed $N$. 
Well below the condensation temperature $\langle N_{\rm ex}\rangle_N$ 
does not depend on $N$, so from Eq. (\ref{fluctNex}) we get 
$\delta N_{\rm ex}=\delta^{\rm c} N_{\rm ex}$ and 
\begin{equation}
\langle\delta^2 N_{\rm ex}\rangle = \langle\delta^2 N_{\rm ex}\rangle^c 
= \langle\delta^2 N_0\rangle^c \,. \label{fluctex}
\end{equation}

\section{Conclusions}
\label{conclusions}

The thermodynamic equivalence between ideal bosons and fermions with 
the same constant density of states $\sigma$ 
\cite{auluck,may,viefers,lee,apostol,crescimanno,paper1} apparently is 
lost below the Bose-Einstein condensation temperature \cite{pathria}. 
On the other hand it was proven that if the Bose and the Fermi systems 
have the same spectrum consisting of nondegenerate, equidistant single 
particle states (like for example particles in a one-dimensional 
harmonic potential), then 
the canonical thermodynamic equivalence between the two systems is 
preserved down to zero temperature 
in the smallest details \cite{crescimanno,paper1}. 
This apparent contradiction is due to the fact that below the condensation 
temperature $T_{\rm c}$, in the Fermi gas the values of intensive 
parameters like the 
chemical potential and also the population of single particle levels in the 
canonical ensemble are 
changed slightly from their corresponding values in the grandcanonical 
ensemble. 

Below $T_{\rm c}$, to the Bose-Einstein condensate in the Bose system it 
corresponds in the Fermi system a degenerate gas of the same number of 
particles, at the lowest part of the spectrum. Because of this correspondence 
the degenerate fermionic 
subsystem is called here the Fermi condensate \cite{paper2}. 
The Fermi condensate is manifested also in the probability distribution 
of the grandcanonical ensemble. One can calculate the probability to have 
$N_0$ degenerate particles (see Eq. 1). Below  $T_{\rm c}$ 
this probability distribution has a maximum for $N_0>0$ and we showed 
numerically and analytically in Section \ref{exact} that the 
grandcanonical average of $N_0$ is the same as the canonical average.

In this paper we did both, analytical and numerical calculations of 
the number of particles in the condensate and in the thermal active 
layer. We calculated also their fluctuations. 
The grandcanonical fluctuation of $N_0$ is almost the same as the 
fluctuation in the canonical ensemble. Although the average values 
$\langle N_0\rangle$ and $\langle N_0\rangle^{\rm c}$ are identical, 
for large values of $\sigma\kb T$ the fluctuations 
$\langle\delta^2 N_0\rangle^{1/2}$ and 
$(\langle\delta^2 N_0\rangle^{\rm c})^{1/2}$ differ by a small, but 
constant value, 0.39. This is due to the extra contribution to 
$\langle\delta^2 N_0\rangle^{1/2}$, given by the grandcanonical fluctuation of 
the total particle number. 

The fermions in the thermally active layer correspond to the 
bosons on the excited energy levels. Canonical and 
grandcanonical averages of $N_{\rm ex}$ are the same. 
Moreover, well below the condensation temperature, where Maxwell 
Deamon's ensemble is applicable \cite{maxwell}, the fluctuation of 
$N_{\rm ex}$ is the same in both, canonical and grandcanonical ensembles 
(\ref{fluctex}).

\end{document}